\documentclass[prd,showpacs,showkeys,nofootinbib,twocolumn]{revtex4-1}

\usepackage{longtable}

\begin{document}

\title{Conservation laws in gravitational theories with general nonminimal coupling}

\author{Yuri N. Obukhov}
\email{yo@thp.uni-koeln.de}
\affiliation{Theoretical Physics Laboratory, Nuclear Safety Institute, 
Russian Academy of Sciences, B.Tulskaya 52, 115191 Moscow, Russia}

\author{Dirk Puetzfeld}
\email{dirk.puetzfeld@zarm.uni-bremen.de}
\homepage{http://puetzfeld.org}
\affiliation{ZARM, University of Bremen, Am Fallturm, 28359 Bremen, Germany} 

\date{ \today}

\begin{abstract}
We use the Lagrange-Noether methods to derive the conservation laws for models in which matter interacts nonminimally with the gravitational field. The nonminimal coupling function can depend arbitrarily on the gravitational field strength. The obtained result generalizes earlier findings. The generalized conservation laws provide the basis for the derivation of the equations of motion for the nonminimally coupled test bodies.
\end{abstract}

\pacs{04.20.Fy; 04.50.Kd; 04.20.Cv}
\keywords{Conservation laws; Noether theorem; Variational methods}

\maketitle

%% 04.20.Fy Variational methods in general relativity
%% 04.50.Kd Gravity modified theories of
%% 04.20.Cv Fundamental problems and general formalism

\section{Introduction}\label{introduction_sec}

Recently modified gravity theories with nonminimal coupling have attracted considerable attention. In such models matter interacts with the gravitational field directly via the explicit dependence of the generalized {\it coupling function} (that replaces the coupling {\it constant}) on the curvature of spacetime. See refs.\ \cite{Bertolami:etal:2007,Mohseni:2009,Mohseni:2010}; the reviews \cite{Schmidt:2007,Straumann:2008,Nojiri:2011} give an outlook and contain references for the further reading. 

It was soon recognized that the nonminimal coupling leads to a modification of the conservation laws of the energy-momentum (for the early analysis see \cite{Koivisto:2006} and \cite{Bertolami:etal:2007}). This is an important observation since the conservation laws underlie the derivation of the equations of motion of the test bodies. As a result, the massive extended bodies and particles are affected by an extra force, as compared to the minimally coupled case
\cite{Puetzfeld:Obukhov:2008:1,Puetzfeld:Obukhov:2013}. Some recent works \cite{Mohseni:2009,Mohseni:2010} reported very complicated modifications of the conservation laws.

Here we carefully analyze the consequences of the general coordinate invariance of the action that describes the most general nonminimal coupling of matter to the
gravitational field strengths. The Lagrange-Noether framework yields the conservation laws which have a remarkably simple structure. Our results generalize \cite{Puetzfeld:Obukhov:2013} and correct \cite{Mohseni:2009,Mohseni:2010} earlier derivations. In particular, recently in \cite{Puetzfeld:Obukhov:2013} we have considered the case when the nonminimal coupling function depends arbitrarily on the 9 parity-even curvature invariants. These belong to the set of the 14 algebraically independent invariants constructed from the components of the Riemann tensor, which characterize a curved spacetime of 4 dimensions \cite{Thomas:1934,Debever:1964,Carminati:1991}. Our newly derived general conservation laws extend the aforementioned results.

Furthermore, our current analysis also covers the case when the material elements have the microstructural properties such as spin. The resulting conservation laws then are suitable for the study of the equations of motion of extended bodies constructed from matter with microstructure coupled nonminimally to the gravitational field. This extends the findings of \cite{Stoeger:Yasskin:1979,Stoeger:Yasskin:1980,Puetzfeld:Obukhov:2007}.  

Our notations and conventions are those of \cite{Hehl:1995}. In particular, the basic geometrical quantities such as the curvature, torsion, etc., are defined as in \cite{Hehl:1995}, and we use the Latin alphabet to label the spacetime coordinate indices. Furthermore, the metric has the signature $(+,-,-,-)$. As a result, our definition of the metrical energy-momentum tensor is different from the definition used in \cite{Bertolami:etal:2007,Nojiri:2011,Puetzfeld:Obukhov:2013}. 

The structure of the paper is as follows: In section \ref{sec_models} we briefly introduce different nonminimal coupling scenarios, and in particular we formulate the maximally extended version of nonminimal gravity, in which the coupling function can depend arbitrarily on the metric, curvature, and torsion of spacetime. The general Lagrange-Noether analysis is developed in section \ref{Noether_sec}, and the results obtained are subsequently applied in Sec.\ \ref{conservation_sec} to the extended nonminimal model, for which we explicitly work out the conservation laws. A further generalization to nonminimally coupled matter with intrinsic moments is considered in Sec.\ \ref{quadrupole}. Finally, our findings are discussed in section \ref{conclusion_sec}.

\section{Formulation of the problem} \label{sec_models}

\subsection{Nonminimal $f(R)$ gravity} \label{subsec_model1}

In \cite{Bertolami:etal:2007,Schmidt:2007,Straumann:2008,Nojiri:2011} an extended version of a so-called $f(R)$ gravity theory was considered. The corresponding interaction Lagrangian was put forward,
\begin{eqnarray}
L =  \left[1+ \lambda f_2\left( R\right) \right] L_{\rm mat}, \label{ansatz_lagrangian}
\end{eqnarray} 
where the nonminimal coupling function $f_2$ depends arbitrarily on the curvature scalar $R$, and $L_{\rm mat}$ is the matter Lagrangian. The nonminimal coupling of matter and gravity is controlled by the constant $\lambda$. 

In contrast to standard general relativity theory, the last term in (\ref{ansatz_lagrangian}) leads to a modification of the conservation law of the metrical energy-momentum tensor of matter defined by $\sqrt{-g}t_{ij} :=  2\delta (\sqrt{-g}L_{\rm mat})/\delta g^{ij}$. It reads
\begin{eqnarray}
\nabla^i t_{ij} = \frac{\lambda f'_2}{1+\lambda f_2} \left( - g_{ij} L_{\rm mat} - t_{ij} \right) \nabla^i R. \label{conservation}
\end{eqnarray}  
Here $f'_2\left(R\right):=df_2\left(R\right)/dR $ denotes a shortcut for derivatives of the unspecified function $f_2\left(R\right)$ of the curvature scalar. The first term in the parentheses on the right-hand side has the different sign, as compared to \cite{Bertolami:etal:2007,Nojiri:2011,Puetzfeld:Obukhov:2013}, due to a different metric signature and the corresponding different definition of the metrical energy-momentum tensor. 

\subsection{Generalized nonminimal gravity} \label{subsec_model2}

In \cite{Puetzfeld:Obukhov:2013} the above model was generalized to
\begin{eqnarray}
L =  F L_{\rm mat}, \label{ansatz_lagrangian_model_2}
\end{eqnarray}
where the nonminimal coupling function $F = F(i_1,\dots, i_9)$ depends on the set of the 9 parity-even invariants constructed from the components of the curvature tensor,
\begin{eqnarray}\label{invariants}
i_1 &=& R^2,\qquad i_2 = R_{ij} R^{ij},\qquad i_3 = R_{ijkl} R^{ijkl},\\
i_4 &=& R_{ij}{}^{kl}R_{kl}{}^{mn}R_{mn}{}^{ij},\label{inv4}\\ 
i_5 &=& R^i{}_jR^j{}_kR^k{}_i,\quad
i_6 = R^i{}_jR^j{}_kR^k{}_lR^l{}_i,\label{inv56}\\
i_7 &=& R^{ij}D_{ij},\quad i_8 = D_{ij} D^{ij},\quad i_9 = D_{ij} D^{jk}R^i{}_k.\label{inv789}
\end{eqnarray}
Here we have denoted $D_{ij} := R_{iklj}R^{kl}$. The set (\ref{invariants})-(\ref{inv789}) is equivalent to the one reported in \cite{Debever:1964,Carminati:1991}, when the Riemann tensor is decomposed in terms of the Weyl and the traceless Ricci tensor. 

Generalized gravity theories with Lagrangians that are functions of the minimal independent set of curvature invariants have recently attracted some attention in the cosmological context; see \cite{Ishak:2009}, for example. 

Without using the Noether theorem, in \cite{Puetzfeld:Obukhov:2013} we demonstrated directly from the field equations that the conservation law for the model (\ref{ansatz_lagrangian_model_2}) reads
\begin{eqnarray}\label{conservation2}
\nabla^i t_{ij} = {\frac 1F} \left( - g_{ij} L_{\rm mat} - t_{ij} \right)\nabla^iF.
\end{eqnarray}  
This result generalizes the conservation law (\ref{conservation}) to the case in which $F = F(i_1,\dots,i_9)$ depends arbitrarily on the complete set of 9 parity-even curvature invariants (\ref{invariants})-(\ref{inv789}), correcting the earlier derivations \cite{Mohseni:2009,Mohseni:2010}. Again notice a different conventional sign, as compared to our previous work \cite{Puetzfeld:Obukhov:2013}. 

\subsection{Maximally extended nonminimal gravity} \label{subsec_model3}

In order to be as general as possible, we consider the matter with microstructure, namely, with spin. An appropriate gravitational model is then the Poincar\'e gauge theory in which the metric tensor $g_{ij}$ is accompanied by the connection $\Gamma_{ki}{}^j$ that is metric-compatible but not necessarily symmetric. The gravitational field strengths are the Riemann-Cartan curvature and the torsion,
\begin{eqnarray}
R_{kli}{}^j &=& \partial_k\Gamma_{li}{}^j - \partial_l\Gamma_{ki}{}^j + \Gamma_{kn}{}^j \Gamma_{li}{}^n - \Gamma_{ln}{}^j\Gamma_{ki}{}^n,\label{curv}\\
T_{kl}{}^i &=& \Gamma_{kl}{}^i - \Gamma_{lk}{}^i.\label{tors}
\end{eqnarray}

Our aim is to study the nonminimal gravity model in which the interaction Lagrangian reads
\begin{eqnarray}
L =  F(g_{ij},R_{kli}{}^j,T_{kl}{}^i) L_{\rm mat}.\label{ansatz_lagrangian_model_3}
\end{eqnarray}
The coupling function $ F(g_{ij},R_{kli}{}^j,T_{kl}{}^i)$ depends arbitrarily on its arguments. In technical terms, $F$ is a function of independent scalar invariants constructed in all possible ways from the components of the curvature and torsion tensors.

\section{Lagrange-Noether analysis}\label{Noether_sec}

The basic ideas and the general scheme, as well as the exhaustive literature, can be found in \cite{Trautman:1962,Kosmann:2011}. We will follow quite closely along the lines of the standard discussion of the Noether theorem.

It is convenient to embed the problem formulated above into a wider framework that deals with a general action 
\begin{equation}
I = \int\,d^4x\,{\cal L}.\label{action}
\end{equation}
The Lagrangian density ${\cal L} = {\cal L}(\Phi^J,\partial_i\Phi^J)$ depends on the set of fields which we collectively denote $\Phi^J = (g_{ij},\Gamma_{ki}{}^j,\psi^A)$. We do not specify the range of the multi-index ${}^J$ at this stage. 

Let us consider arbitrary infinitesimal transformation of the spacetime coordinates and the matter fields
\begin{eqnarray}
x^i &\longrightarrow& x'^i (x) = x^i + \delta x^i,\label{dx}\\
\Phi^J(x) &\longrightarrow& \Phi'^J(x') = \Phi^J(x) + \delta\Phi^J(x).
\label{dP}
\end{eqnarray} 
Within the present context it is not important whether this is a symmetry transformation under the action of any specific group. The total variation (\ref{dP}) is a result of the change of the form of the functions and of the change induced by the transformation of the spacetime coordinates (\ref{dx}). In order to distinguish the two pieces in the field transformation, it is convenient to introduce the {\it substantial variation},
\begin{equation}\label{sub}
\overline{\delta}\Phi^J := \Phi'^J(x) - \Phi^J(x) = \delta\Phi^J - \delta x^k\partial_k\Phi^J.
\end{equation}
By definition, the substantial variation commutes with the partial derivative, $\overline{\delta}\partial_i = \partial_i\overline{\delta}$. 

We need the total variation of the action,
\begin{equation}
\delta I = \int \left[d^4x\,\delta{\cal L} + \delta(d^4x)\,{\cal L}\right].
\end{equation}
A standard derivation shows that under the action of the transformation (\ref{dx})-(\ref{dP}), the total variation reads
\begin{equation}
\delta I = \int d^4x \left[ {\frac {\delta {\cal L}}{\delta\Phi^J}} \,\overline{\delta}\Phi^J + \partial_i\left({\cal L}\,\delta x^i + {\frac {\partial {\cal L}}{\partial(\partial_i\Phi^J)}}\,\overline{\delta} \Phi^J \right)\right]. \label{master}
\end{equation}
Here the variational derivative is defined, as usual, by
\begin{equation}\label{var}
{\frac {\delta {\cal L}}{\delta\Phi^J}} := {\frac {\partial {\cal L}}{\partial\Phi^J}} - \partial_i\left({\frac {\partial {\cal L}}{\partial(\partial_i\Phi^J)}}\right).
\end{equation}

\subsection{General coordinate invariance}

Under the general coordinate transformations, we have $x^i\rightarrow x^i + \delta x^i$, $g_{ij}\rightarrow g_{ij} + \delta g_{ij}$, $\Gamma_{ki}{}^j\rightarrow \Gamma_{ki}{}^j + \delta\Gamma_{ki}{}^j$, and $\psi^A \rightarrow \psi^A + \delta\psi^A$ with
\begin{eqnarray}
\delta x^i &=& \xi^i(x),\label{dex}\\ \label{dgij}
\delta g_{ij} &=& -\,(\partial_i\xi^k)\,g_{kj} - (\partial_j\xi^k)\,g_{ik},\\
\delta\psi^A&=&-\,(\partial_i\xi^j)\,(\sigma^A{}_B)_j{}^i\,\psi^B, \label{dpsiA}\\
\delta\Gamma_{ki}{}^j &=&  -\,(\partial_k\xi^l)\,\Gamma_{li}{}^j - (\partial_i\xi^l)\,\Gamma_{kl}{}^j \nonumber\\ 
&& + \,(\partial_l\xi^j)\,\Gamma_{ki}{}^l - \partial^2_{ki}\xi^j.\label{dG}
\end{eqnarray}
Here $(\sigma^A{}_B)_j{}^i$ are generators of the general coordinate transformations which satisfy the commutation relations
\begin{eqnarray}
(\sigma^A{}_C)_j{}^i(\sigma^C{}_B)_l{}^k - (\sigma^A{}_C)_l{}^k (\sigma^C{}_B)_j{}^i\nonumber\\ 
= (\sigma^A{}_B)_l{}^i\,\delta^k_j - (\sigma^A{}_B)_j{}^k \,\delta^i_l.\label{comms}
\end{eqnarray}

Substituting (\ref{dex})-(\ref{dG}) into (\ref{master}), and making use of the substantial derivative definition (\ref{sub}), we find
\begin{eqnarray}
\delta I &=& -\,\int d^4x \biggl[\xi^k\,\Omega_k + (\partial_i\xi^k)\,\Omega_k{}^i\nonumber\\
&& +\,(\partial^2_{ij}\xi^k)\,\Omega_k{}^{ij} + (\partial^3_{ijn}\xi^k)\,\Omega_k{}^{ijn}\biggr],
\label{masterG} 
\end{eqnarray}
where explicitly
\begin{eqnarray}
\Omega_k &=& {\frac {\partial {\cal L}}{\partial g_{ij}}}\,\partial_kg_{ij} + {\frac {\delta {\cal L}}{\delta\psi^A}}\,\partial_k\psi^A \nonumber\\
&& + \,\partial_i\left({\frac {\partial {\cal L}}{\partial\partial_i\psi^A}} \,\partial_k\psi^A - \delta^i_k{\cal L} \right)\nonumber\\
&& + \,{\frac {\partial {\cal L}}{\partial \Gamma_{ln}{}^m}}\,\partial_k\Gamma_{ln}{}^m + {\frac {\partial {\cal L}}{\partial \partial_i\Gamma_{ln}{}^m}} \,\partial_k\partial_i\Gamma_{ln}{}^m,\label{Om1}\\
\Omega_k{}^i &=& 2{\frac {\partial {\cal L}}{\partial g_{ij}}}\,g_{kj} + {\frac {\delta {\cal L}}{\delta\psi^A}}\,(\sigma^A{}_B)_k{}^i\,\psi^B \nonumber\\ 
&& + {\frac {\partial {\cal L}}{\partial\partial_i\psi^A}}\partial_k\psi^A - \delta^i_k{\cal L} + \partial_j\!\left(\!{\frac {\partial {\cal L}}{\partial\partial_j\psi^A}}(\sigma^A{}_B)_k{}^i\psi^B\!\right)\!\nonumber\\
&& + \,{\frac {\partial {\cal L}}{\partial \Gamma_{li}{}^j}}\,\Gamma_{lk}{}^j + {\frac {\partial {\cal L}}{\partial \Gamma_{il}{}^j}}\,\Gamma_{kl}{}^j - {\frac {\partial {\cal L}}{\partial \Gamma_{lj}{}^k}}\,\Gamma_{lj}{}^i\nonumber\\
&& + \,{\frac {\partial {\cal L}}{\partial \partial_i\Gamma_{ln}{}^m}}\,\partial_k \Gamma_{nl}{}^m + {\frac {\partial {\cal L}}{\partial \partial_n\Gamma_{il}{}^m}} \,\partial_n\Gamma_{kl}{}^m \nonumber\\
&& + \,{\frac {\partial {\cal L}}{\partial \partial_n\Gamma_{li}{}^m}}\,\partial_n \Gamma_{lk}{}^m - {\frac {\partial {\cal L}}{\partial \partial_n\Gamma_{lm}{}^k}}
\,\partial_n\Gamma_{lm}{}^i ,\label{Om2}\\
\Omega_k{}^{ij} &=& {\frac {\partial {\cal L}}{\partial\partial_{(i}\psi^A}} (\sigma^A{}_B)_k{}^{j)}\psi^B + {\frac {\partial {\cal L}}{\partial \Gamma_{(ij)}{}^k}}
+ {\frac {\partial {\cal L}}{\partial \partial_{(i}\Gamma_{j)l}{}^m}}\Gamma_{kl}{}^m \nonumber\\
&& + \,{\frac {\partial {\cal L}}{\partial \partial_{(i}\Gamma_{|l|j)}{}^m}}\,\Gamma_{lk}{}^m - {\frac {\partial {\cal L}}{\partial \partial_{(i}\Gamma_{|ln|}{}^k}}\,\Gamma_{ln}{}^{j)}. \label{Om3}\\ \label{Om4}
\Omega_k{}^{ijn} &=& {\frac {\partial {\cal L}}{\partial \partial_{(n}\Gamma_{ij)}{}^k}}.
\end{eqnarray}
If the action is invariant under the general coordinate transformations, $\delta I = 0$, in view of the arbitrariness of the function $\xi^i$ and its derivatives, we find the set of the four Noether identities:
\begin{equation}\label{NoeG}
\Omega_k = 0,\quad \Omega_k{}^i = 0,\quad \Omega_k{}^{ij} = 0, \quad \Omega_k{}^{ijn} = 0.
\end{equation}

General coordinate invariance is a natural consequence of the fact that the action (\ref{action}) and the Lagrangian ${\cal L}$ are constructed only from covariant objects. Namely, ${\cal L} = {\cal L}(\psi^A, \nabla_i\psi^A, g_{ij}, R_{kli}{}^j, T_{kl}{}^i)$ is a function of the metric, the curvature (\ref{curv}), the torsion (\ref{tors}), the matter field, and its {\it covariant derivative} 
\begin{equation}\label{Dpsi}
\nabla_k\psi^A = \partial_k\psi^A -\Gamma_{ki}{}^j\,(\sigma^A{}_B)_j{}^i\,\psi^B.
\end{equation}
Denoting 
\begin{equation}
\rho^{ijk}{}_l := {\frac {\partial {\cal L}}{\partial R_{ijk}{}^l}},\qquad
\sigma^{ij}{}_k := {\frac {\partial {\cal L}}{\partial T_{ij}{}^k}},\label{dLRT}
\end{equation}
we find for the derivatives of the Lagrangian
\begin{eqnarray}
{\frac {\partial {\cal L}}{\partial \Gamma_{ij}{}^k}} &=& -\,{\frac {\partial {\cal L}} {\partial \nabla_i\psi^A}}(\sigma^A{}_B)_k{}^j\,\psi^B + 2\sigma^{ij}{}_k\nonumber\\
&& + 2\rho^{inl}{}_k\Gamma_{nl}{}^j + 2\rho^{nij}{}_l\Gamma_{nk}{}^l,\label{dLG}\\
{\frac {\partial {\cal L}}{\partial \partial_i\Gamma_{jk}{}^l}} &=& 2\rho^{ijk}{}_l.\label{dLdG}
\end{eqnarray}
As a result, we straightforwardly verify that $\Omega_k{}^{ij} = 0$ and $\Omega_k{}^{ijn} = 0$ are indeed satisfied identically. 

Using (\ref{dLG}) and (\ref{dLdG}), we then recast the two remaining Noether identities (\ref{Om1}) and (\ref{Om2}) into
\begin{eqnarray}
\Omega_k &=& {\frac {\partial {\cal L}}{\partial g_{ij}}}\,\partial_kg_{ij} + {\frac {\delta {\cal L}}{\delta\psi^A}}\,\partial_k\psi^A \nonumber\\
&& + \,\partial_i\!\left(\!{\frac {\partial {\cal L}}{\partial\nabla_i\psi^A}} \,\nabla_k\psi^A - \delta^i_k{\cal L} \!\right)\!\nonumber\\
&& +\,\widehat{\nabla}{}_j\!\left(\!{\frac {\partial {\cal L}}{\partial\nabla_j\psi^A}} \,(\sigma^A{}_B)_m{}^n\,\psi^B\!\right)\!\Gamma_{kn}{}^m\nonumber\\ 
&& + \,{\frac {\partial {\cal L}}{\partial\nabla_l\psi^A}}\,(\sigma^A{}_B)_m{}^n \,\psi^B\,R_{lkn}{}^m \nonumber\\
&& + \,\rho^{iln}{}_m\partial_kR_{iln}{}^m + \sigma^{ln}{}_m\partial_kT_{ln}{}^m =0,\label{Om1a}\\
\Omega_k{}^i &=& 2{\frac {\partial {\cal L}}{\partial g_{ij}}}\,g_{kj} + {\frac {\delta {\cal L}}{\delta\psi^A}}\,(\sigma^A{}_B)_k{}^i\,\psi^B \nonumber\\ 
&& + {\frac {\partial {\cal L}}{\partial\nabla_i\psi^A}}\nabla_k\psi^A - \delta^i_k{\cal L} 
+ \widehat{\nabla}{}_j\!\!\left(\!{\frac {\partial {\cal L}}{\partial\nabla_j\psi^A}}
(\sigma^A{}_B)_k{}^i\psi^B\!\right)\!\nonumber\\
&& + \,2\rho^{iln}{}_mR_{kln}{}^m + \rho^{lni}{}_mR_{lnk}{}^m - \rho^{lnm}{}_kR_{lnm}{}^i\nonumber\\
&& + \,2\sigma^{il}{}_nT_{kl}{}^n - \sigma^{ln}{}_kT_{ln}{}^i =0.\label{Om2a}
\end{eqnarray}
Here we introduced the covariant derivative for an arbitrary tensor density ${\cal A}^n{}_{i\dots}{}^{j\dots}$
\begin{equation}
\widehat{\nabla}{}_n{\cal A}^n{}_{i\dots}{}^{j\dots} = \partial_n{\cal A}^n{}_{i\dots}{}^{j\dots} + \Gamma_{nl}{}^j{\cal A}^n{}_{i\dots}{}^{l\dots} - \Gamma_{ni}{}^l
{\cal A}^n{}_{l\dots}{}^{j\dots},\label{dA}
\end{equation}
which produces a tensor density of the same weight. In particular, notice that the variational derivative (\ref{var}) of the matter field, identically rewritten as
\begin{equation}\label{covar}
{\frac {\delta {\cal L}}{\delta\psi^A}} = {\frac {\partial {\cal L}}{\partial\psi^A}} - \widehat{\nabla}{}_j\left({\frac {\partial {\cal L}}{\partial\nabla_j\psi^A}}\right),
\end{equation}
is a covariant tensor density. 

The Noether identity (\ref{Om2a}) is a covariant relation. In contrast, (\ref{Om1a}) is apparently noncovariant. However, this can be easily repaired by replacing $\Omega_k = 0$ with an equivalent covariant Noether identity: $\overline{\Omega}{}_k = \Omega_k -  \Gamma_{kn}{}^m\Omega_m{}^n = 0$. Explicitly, we find
\begin{eqnarray}
\overline{\Omega}{}_k &=& {\frac {\delta {\cal L}}{\delta\psi^A}}\,\nabla_k\psi^A + \widehat{\nabla}{}_i\!\left(\!{\frac {\partial {\cal L}}{\partial\nabla_i\psi^A}} \,\nabla_k\psi^A - \delta^i_k{\cal L} \!\right)\!\nonumber\\
&& -\,\left({\frac {\partial {\cal L}}{\partial\nabla_i\psi^A}} \,\nabla_l\psi^A - \delta^i_l{\cal L}\right) T_{ki}{}^l\nonumber\\ 
&& +\,{\frac {\partial {\cal L}}{\partial\nabla_l\psi^A}}\,(\sigma^A{}_B)_m{}^n \,\psi^B\,R_{lkn}{}^m \nonumber\\
&& + \,\rho^{iln}{}_m\nabla_kR_{iln}{}^m + \sigma^{ln}{}_m\nabla_kT_{ln}{}^m =0.\label{Om1b}
\end{eqnarray}

On-shell, i.e., assuming that the matter field satisfies the field equations ${\delta {\cal L}}/{\delta\psi^A} = 0$, the Noether identities (\ref{Om2a}) and (\ref{Om1b}) reduce to the {\it conservation laws}  for the energy-momentum and spin.

More exactly, (\ref{Om1b}) gives rise to the conservation law of the energy and the momentum, in which the divergence of the canonical energy-momentum tensor is balanced by the Lorentz-type forces of Mathisson-Papapetrou (second and third lines); it is worthwhile to notice the appearance of the {\it quadrupole-type} terms displayed in the last line. 

Equation (\ref{Om2a}) contains a relation between the canonical and metrical energy-momentum tensors and the conservation law of spin. In the next section we turn to the discussion of the general nonminimal coupling models.

\section{Conservation laws in models with nonminimal coupling}\label{conservation_sec}

The results obtained in the previous section are applicable to {\it any} theory in which the Lagrangian depends arbitrarily on the matter field and the gravitational field strengths. Now we specialize to the class of models described by (\ref{ansatz_lagrangian_model_3}).

\subsection{Identities for the nonminimal coupling function}\label{Fidentities_subsec}

As a preliminary step, let us derive the identities which are satisfied for the nonminimal coupling function $F = F(g_{ij},R_{kli}{}^j,T_{kl}{}^i)$. For this, we apply the above Lagrange-Noether machinery to the auxiliary Lagrangian density ${\cal L}_0 = \sqrt{-g}\,F$. This quantity does not depend on the matter fields, and both (\ref{Om2a}) and (\ref{Om1b}) are considerably simplified. In particular, we have
\begin{equation}
{\frac {\partial {\cal L}_0}{\partial g_{ij}}} = \sqrt{-g}\left({\frac 12}Fg^{ij} + F^{ij} \right),\qquad F^{ij} := {\frac {\partial F}{\partial g_{ij}}}.\label{dFg}
\end{equation}
Then we immediately see that (\ref{Om2a}) and (\ref{Om1b}) reduce to
\begin{eqnarray}
\nabla_k F &=& {\stackrel 0 \rho}{}^{iln}{}_m\nabla_kR_{iln}{}^m + {\stackrel 0 \sigma}{}^{ln}{}_m\nabla_kT_{ln}{}^m,\label{F1}\\
2F_k{}^i &=& - \,2{\stackrel 0 \rho}{}^{iln}{}_mR_{kln}{}^m - {\stackrel 0 \rho}{}^{lni}{}_m R_{lnk}{}^m + {\stackrel 0 \rho}{}^{lnm}{}_kR_{lnm}{}^i\nonumber\\
&& -\,2{\stackrel 0 \sigma}{}^{il}{}_nT_{kl}{}^n + {\stackrel 0 \sigma}{}^{ln}{}_k T_{ln}{}^i.\label{F2}
\end{eqnarray}
Here we denoted
\begin{equation}\label{dFRT}
{\stackrel 0 \rho}{}^{ijk}{}_l := {\frac {\partial F}{\partial R_{ijk}{}^l}},\qquad
{\stackrel 0 \sigma}{}^{ij}{}_k := {\frac {\partial F}{\partial T_{ij}{}^k}}.
\end{equation}
The identity (\ref{F1}) is naturally interpreted as a generally covariant generalization of the chain differentiation rule.

It should be stressed that (\ref{F1}) and (\ref{F2}) are the true identities, they are satisfied for any function $F(g_{ij},R_{kli}{}^j,T_{kl}{}^i)$ irrespectively of the field equations that can be derived from the corresponding action. 

\subsection{Conservation laws}\label{Conslaws_subsec}

Now we are in a position to derive the conservation laws for the general nonminimal coupling model (\ref{ansatz_lagrangian_model_3}), and thus we have to consider the Lagrangian density
\begin{equation}
{\cal L} = \sqrt{-g}FL_{\rm mat}.\label{Lnon}
\end{equation}
As before, $F = F(g_{ij},R_{kli}{}^j,T_{kl}{}^i)$ is an arbitrary function of its arguments, whereas the matter Lagrangian $L_{\rm mat} = L_{\rm mat}(\psi^A, \nabla_i\psi^A, g_{ij})$ has the usual form established from the minimal coupling principle. 

In a standard way, the matter is characterized by the canonical energy-momentum tensor,
\begin{equation}
\Sigma_k{}^i = {\frac {\partial {L_{\rm mat}}}{\partial\nabla_i\psi^A}} \,\nabla_k\psi^A - \delta^i_kL_{\rm mat},\label{emcan}
\end{equation}
the canonical spin tensor,
\begin{equation}
\tau^n{}_k{}^i = -{\frac {\partial {L_{\rm mat}}}{\partial\nabla_i\psi^A}} \,(\sigma^A{}_B)_k{}^n \psi^B,\label{spin}
\end{equation}
and the metrical energy-momentum tensor
\begin{equation}\label{emmet}
t_{ij} = {\frac 2{\sqrt{-g}}}\,{\frac {\partial {(\sqrt{-g}L_{\rm mat})}}{\partial g^{ij}}}.
\end{equation}

In view of the product structure of the Lagrangian (\ref{Lnon}), the derivatives are easily evaluated, and the conservation laws (\ref{Om2a}) and (\ref{Om1b}) reduce to
\begin{eqnarray}
&&- Ft_k{}^i - {\stackrel * \nabla}{}_n\left(F\tau^i{}_k{}^n\right) + F\Sigma_k{}^i\nonumber\\
&&+ \Bigl[2F_k{}^i + 2{\stackrel 0 \rho}{}^{iln}{}_mR_{kln}{}^m + {\stackrel 0 \rho}{}^{lni}{}_m R_{lnk}{}^m - {\stackrel 0 \rho}{}^{lnm}{}_kR_{lnm}{}^i\nonumber\\
&& + 2{\stackrel 0 \sigma}{}^{il}{}_nT_{kl}{}^n - {\stackrel 0 \sigma}{}^{ln}{}_k T_{ln}{}^i\Bigr]L_{\rm mat} = 0,\label{cons1a}\\
&& {\stackrel * \nabla}{}_i\left(F\Sigma_k{}^i\right) - F\Sigma_l{}^i T_{ki}{}^l + F\tau^m{}_n{}^l R_{klm}{}^n \nonumber\\
&& + \Bigl[{\stackrel 0 \rho}{}^{iln}{}_m\nabla_kR_{iln}{}^m + {\stackrel 0 \sigma}{}^{ln}{}_m\nabla_kT_{ln}{}^m\Bigr]L_{\rm mat} = 0.\label{cons2a}
\end{eqnarray}
Here the so-called modified covariant derivative is defined as usual by
\begin{equation}
{\stackrel * \nabla}{}_i = \nabla_i - T_{ki}{}^k.\label{dstar}
\end{equation}
It replaces the derivative (\ref{dA}) when we pass from the tensor densities to the true tensors in the Riemann-Cartan spacetime.

After we take into account the identities (\ref{F1}) and (\ref{F2}), the conservation laws (\ref{cons1a}) and (\ref{cons2a}) are brought to the final form:
\begin{eqnarray}\label{cons1b}
F\Sigma_k{}^i &=& Ft_k{}^i + {\stackrel * \nabla}{}_n\left(F\tau^i{}_k{}^n\right),\\
{\stackrel * \nabla}{}_i\left(F\Sigma_k{}^i\right) &=& F\Sigma_l{}^i T_{ki}{}^l - F\tau^m{}_n{}^l R_{klm}{}^n\nonumber\\ 
&& -\, L_{\rm mat}\nabla_kF.\label{cons2b}
\end{eqnarray}
Lowering the index in (\ref{cons1b}) and antisymmetrizing, we derive the conservation law for the spin
\begin{equation}
F\Sigma_{[ij]} + {\stackrel * \nabla}{}_n\left(F\tau_{[ij]}{}^n\right) = 0.
\end{equation}
This is a generalization of the usual conservation law of the total angular momentum for the case of nonminimal coupling. 

\subsection{Purely Riemannian theory}\label{Riemannian_subsec}

Our results contain the Riemannian theory as a special case. Suppose the torsion is absent $T_{ij}{}^k = 0$. Then for usual matter without microstructure (spinless matter with $\tau^m{}_n{}^i = 0$) the canonical and the metrical energy-momentum tensors coincide, $\Sigma_k{}^i = t_k{}^i$. As a result, the conservation law (\ref{cons2b}) reduces to
\begin{equation}
\nabla_it_k{}^i = {\frac 1F}\left(- L_{\rm mat}\delta_k^i -t_k{}^i\right)\nabla_iF.\label{consF}
\end{equation}
It is remarkable that we are able to generalize the earlier result (\ref{conservation2}) to the case when the nonminimal coupling function $F$ depends not just on the minimal set of the curvature invariants but is actually an absolutely arbitrary scalar function of the curvature tensor.

\section{Further generalization: matter with intrinsic moments}\label{quadrupole}

Our formalism allows to consider also the case when matter couples to the gravitational field strengths not just through an $F$-factor in front of the Lagrangian but directly via Pauli-type interaction terms in $L_{\rm mat}$,
\begin{equation}
I^{klm}{}_n(\psi^A, g_{ij})R_{klm}{}^n + J^{kl}{}_n(\psi^A, g_{ij})T_{kl}{}^n.\label{IJ}
\end{equation}
In Maxwell's electrodynamics, similar terms describe the interaction of the electromagnetic field to the anomalous magnetic and/or electric dipole moments. For the Dirac spinor matter \cite{Obukhov:1998,Hehl:etal:1998}, the Pauli-type quantities $I^{klm}{}_n(\psi^A, g_{ij})$ and $J^{kl}{}_n(\psi^A, g_{ij})$ are interpreted as the (Lorentz and translational, respectively) ``gravitational moments'' that arise from the Gordon decomposition of the dynamical currents.

Then for a Lagrangian density ${\cal L} = \sqrt{-g}L_{\rm mat}$ with $L_{\rm mat}$ that contains Pauli-type terms (\ref{IJ}), we find the derivatives (\ref{dLRT})
\begin{equation}\label{rsIJ}
\rho^{klm}{}_n = \sqrt{-g}\,I^{klm}{}_n,\qquad \sigma^{kl}{}_n = \sqrt{-g}\,J^{kl}{}_n.
\end{equation}
As a result, the Noether identities (\ref{Om2a}) and (\ref{Om1b}) yield the on-shell conservation laws,
\begin{eqnarray}
\Sigma_k{}^i &=& t_k{}^i + {\stackrel * \nabla}{}_n\tau^i{}_k{}^n - 2J^{il}{}_nT_{kl}{}^n + J^{ln}{}_kT_{ln}{}^i \nonumber\\
&& - \,2I^{ilnm}R_{klnm} - 2I^{lnm[i}R_{|lnm|k]},\label{q1a}\\
{\stackrel * \nabla}{}_i\Sigma_k{}^i &=& \Sigma_l{}^i T_{ki}{}^l - \tau^m{}_n{}^l R_{klm}{}^n\nonumber\\ 
&& - \,I^{iln}{}_m\nabla_kR_{iln}{}^m - J^{ln}{}_m\nabla_kT_{ln}{}^m.\label{q2a}
\end{eqnarray}

In the purely Riemannian case of General Relativity, the torsion vanishes and the system (\ref{q1a})-(\ref{q2a}) reduces to
\begin{eqnarray}
\nabla_n\tau_{[ik]}{}^n &=& -\,\Sigma_{[ik]} + 4I_{[i}{}^{lnm}R_{k]lnm},\label{q1b}\\
\nabla_i\Sigma_k{}^i &=&  - \tau^m{}_n{}^l R_{klm}{}^n - I^{ilnm}\nabla_kR_{ilnm}.\label{q2b}
\end{eqnarray}
We displayed only the antisymmetric part in (\ref{q1b}), whereas the symmetric equation describes the relation between the metrical and canonical energy-momentum tensors. When deriving (\ref{q1b}), we took into account that in view of the contraction in (\ref{IJ}), we have the symmetry properties 
\begin{equation}
I^{ijkl} = I^{[ij]kl} = I^{ij[kl]} = I^{klij}.\label{Isym} 
\end{equation}

The form of the system of conservation laws (\ref{q1b})-(\ref{q2b}) is very close to Dixon's equations describing the dynamics of material body with the dipole and quadrupole moments. However, it is important to stress that in contrast to Dixon's {\it integrated} moments of usual structureless matter, $\tau_{[ik]}{}^n$ and $I^{ilnm}$ are the {\it intrinsic} spin and quadrupole moments of matter with microstructure. The above conservation laws can also be viewed as a direct generalization of the ones for spinning particles and polarized media given in \cite{Bailey:Israel:1975}.

The explicit equations of motion of such a matter using multipolar expansion techniques will be discussed elsewhere. 

\section{Conclusion}\label{conclusion_sec}

We have obtained the conservation laws (\ref{cons1b})-(\ref{consF}) for the general theory (\ref{ansatz_lagrangian_model_3}) of matter that interacts nonminimally with gravity via the coupling function $F$, which can depend arbitrarily on the gravitational field strengths (\ref{curv}), (\ref{tors}). In a certain sense, this situation is similar to the scalar-tensor type theory \cite{Fujii:Maeda:2003}, where the gravitational coupling constant is replaced by a scalar field that depends on time and spatial coordinates. 

In our study we assumed that the connection is metric-compatible, that is, $Q_{kij} = - \nabla_kg_{ij} = 0$. However, it is straightforward to generalize all derivations to the geometries with nontrivial nonmetricity $Q_{ijk}$. 

We demonstrate that an even further generalization of the gravitational theories with nonminimal coupling is possible by allowing for the direct interaction via Pauli-type gravitational moments, thus extending the earlier results of \cite{Bailey:Israel:1975}.

The results obtained in this work should form the basis for a reanalysis of the equations of motion of the material bodies with microstructure, thus generalizing the previous work \cite{Stoeger:Yasskin:1979,Stoeger:Yasskin:1980,Puetzfeld:Obukhov:2007,Puetzfeld:Obukhov:2008:1} to the general framework with the nonminimal coupling. In particular, this will allow us to extend the discussion on probing possible post-Riemannian spacetime structures by means of the Gravity Probe B mission. 

\section*{Acknowledgements}
This work was supported by the Deutsche Forschungsgemeinschaft (DFG) through the grant LA-905/8-1 (D.P.). 

\bibliographystyle{unsrtnat}
\bibliography{consnon}

\begin{thebibliography}{23}
\providecommand{\natexlab}[1]{#1}
\providecommand{\url}[1]{\texttt{#1}}
\expandafter\ifx\csname urlstyle\endcsname\relax
  \providecommand{\doi}[1]{doi: #1}\else
  \providecommand{\doi}{doi: \begingroup \urlstyle{rm}\Url}\fi

\bibitem[{Bertolami} et~al.(2007){Bertolami}, {B\"ohmer}, {Harko}, and
  {Lobo}]{Bertolami:etal:2007}
O.~{Bertolami}, C.~G. {B\"ohmer}, T.~{Harko}, and F.~S.~N. {Lobo}.
\newblock {Extra force in $f(R)$ modified theories of gravity}.
\newblock \emph{Phys. Rev. D.}, 75:\penalty0 104016, 2007.

\bibitem[{Mohseni}(2009)]{Mohseni:2009}
M.~{Mohseni}.
\newblock {Non-geodesic motion in $f(G)$ gravity with non-minimal coupling}.
\newblock \emph{Phys. Lett. B}, 682:\penalty0 89, 2009.

\bibitem[{Mohseni}(2010)]{Mohseni:2010}
M.~{Mohseni}.
\newblock {Motion of pole-dipole and quadrupole particles in nonminimally
  coupled theories of gravity}.
\newblock \emph{Phys. Rev. D}, 81:\penalty0 124039, 2010.

\bibitem[{Schmidt}(2007)]{Schmidt:2007}
H.~J. {Schmidt}.
\newblock {Fourth order gravity: equations, history, and applications to
  cosmology}.
\newblock \emph{Int. J. Geom. Meth. Mod. Phys.}, 4:\penalty0 209, 2007.

\bibitem[{Straumann}(2008)]{Straumann:2008}
N.~{Straumann}.
\newblock {Problems with Modified Theories of Gravity, as Alternatives to Dark
  Energy}.
\newblock 2008.
\newblock URL \url{arXiv:0809.5148v1 [gr-qc]}.

\bibitem[{Nojiri} and {Odintsov}(2011)]{Nojiri:2011}
S.~{Nojiri} and S.~D. {Odintsov}.
\newblock {Unified cosmic history in modified gravity: from $F(R)$ theory to
  Lorentz non-invariant models}.
\newblock \emph{Phys. Rep.}, 505:\penalty0 59, 2011.

\bibitem[Koivisto(2006)]{Koivisto:2006}
T.~Koivisto.
\newblock {A note on covariant conservation of energy–momentum in modified
  gravities}.
\newblock \emph{Class. Quantum Grav.}, 23:\penalty0 4289, 2006.

\bibitem[{Puetzfeld} and {Obukhov}(2008)]{Puetzfeld:Obukhov:2008:1}
D.~{Puetzfeld} and Yu.~N. {Obukhov}.
\newblock {Motion of test bodies in theories with nonminimal coupling}.
\newblock \emph{Phys. Rev. D}, 78:\penalty0 121501, 2008.

\bibitem[{Puetzfeld} and {Obukhov}(2013)]{Puetzfeld:Obukhov:2013}
D.~{Puetzfeld} and Yu.~N. {Obukhov}.
\newblock {Covariant equations of motion for test bodies in gravitational
  theories with general nonminimal coupling}.
\newblock \emph{Phys. Rev. D}, 87:\penalty0 044045, 2013.

\bibitem[Thomas(1934)]{Thomas:1934}
T.~Y. Thomas.
\newblock \emph{{The differential invariants of generalized spaces}}.
\newblock Cambridge University Press, Cambridge, 1934.

\bibitem[Debever(1964)]{Debever:1964}
R.~Debever.
\newblock {Le rayonnement gravitationnel. Le tenseur de Riemann en relativit\'e
  g\'en\'erale}.
\newblock \emph{Cahiers Phys.}, 168:\penalty0 303, 1964.

\bibitem[{Carminati} and {McLenaghan}(1991)]{Carminati:1991}
J.~{Carminati} and R.~{McLenaghan}.
\newblock {Algebraic invariants of the Riemann tensor in a four-dimensional
  Lorentzian space}.
\newblock \emph{J. Math. Phys.}, 32:\penalty0 3135, 1991.

\bibitem[{Stoeger} and {Yasskin}(1979)]{Stoeger:Yasskin:1979}
W.~R. {Stoeger} and P.~B. {Yasskin}.
\newblock {Can a macroscopic gyroscope feel torsion?}
\newblock \emph{Gen. Rel. Grav.}, 11:\penalty0 427, 1979.

\bibitem[{Yasskin} and {Stoeger}(1980)]{Stoeger:Yasskin:1980}
P.~B. {Yasskin} and W.~R. {Stoeger}.
\newblock {Propagation equations for test bodies with spin and rotation in
  theories of gravity with torsion}.
\newblock \emph{Phys. Rev. D}, 21:\penalty0 2081, 1980.

\bibitem[{Puetzfeld} and {Obukhov}(2007)]{Puetzfeld:Obukhov:2007}
D.~{Puetzfeld} and Yu.~N. {Obukhov}.
\newblock {Propagation equations for deformable test bodies with microstructure
  in extended theories of gravity}.
\newblock \emph{Phys. Rev. D.}, 76:\penalty0 084025, 2007.

\bibitem[{Hehl} et~al.(1995){Hehl}, {McCrea}, {Mielke}, and
  {Ne'eman}]{Hehl:1995}
F.~W. {Hehl}, J.~D. {McCrea}, E.~W. {Mielke}, and Y.~{Ne'eman}.
\newblock {Metric-affine gauge theory of gravity: Field equations, Noether
  identities, world spinors, and breaking of dilation invariance}.
\newblock \emph{Phys. Rep.}, 258:\penalty0 1, 1995.

\bibitem[{Ishak} and {Moldenhauer}(2009)]{Ishak:2009}
M.~{Ishak} and J.~{Moldenhauer}.
\newblock {A minimal set of invariants as a systematic approach to higher order
  gravity models}.
\newblock \emph{JCAP}, 01:\penalty0 024, 2009.

\bibitem[{Trautman}(1962)]{Trautman:1962}
A.~{Trautman}.
\newblock {Conservation laws in general relativity}.
\newblock \emph{in: ``Gravitation: An introduction to current research'', L.
  Witten (ed.), John Wiley \& Sons, New York}, page 169, 1962.

\bibitem[{Kosmann-Schwarzbach}(2011)]{Kosmann:2011}
Y.~{Kosmann-Schwarzbach}.
\newblock \emph{{The Noether Theorems. Invariance and Conservation Laws in the
  Twentieth Century}}.
\newblock Springer, New York, 2011.

\bibitem[{Obukhov}(1998)]{Obukhov:1998}
Yu.~N. {Obukhov}.
\newblock {The gravitational moments of a Dirac particle}.
\newblock \emph{Acta Phys. Pol. B}, 29:\penalty0 1131, 1998.

\bibitem[{Hehl} et~al.(1998){Hehl}, {Mac\'{i}as}, {Mielke}, and
  {Obukhov}]{Hehl:etal:1998}
F.W. {Hehl}, A.~{Mac\'{i}as}, E.~W. {Mielke}, and Yu.~N. {Obukhov}.
\newblock {On the structure of the energy-momentum and the spin currents in
  Dirac's electron theory}.
\newblock \emph{in: ``On Einstein's Path -- Essays in honor of Engelbert
  Schucking'', A. Harvey (ed.), Springer, New York}, page 257, 1998.

\bibitem[{Bailey} and {Israel}(1975)]{Bailey:Israel:1975}
I.~{Bailey} and W.~{Israel}.
\newblock {Lagrangian dynamics of spinning particles and polarized media in
  General Relativity}.
\newblock \emph{Comm. Math. Phys.}, 42:\penalty0 65, 1975.

\bibitem[{Fujii} and {Maeda}(2003)]{Fujii:Maeda:2003}
Y.~{Fujii} and K.~{Maeda}.
\newblock \emph{{The scalar-tensor theory of gravitation}}.
\newblock Cambridge University Press, Cambridge, 2003.

\end{thebibliography}
\end{document}